\documentclass[prl,preprint,showpacs,showkeys,preprintnumbers,amsmath,amssymb]{revtex4-1}

\usepackage{graphicx}
\usepackage{dcolumn}
\usepackage{bm}

\begin{document}

\title{Dissipative Effects on Quantum Sticking}

\author{Yanting Zhang}

\author{Dennis P. Clougherty}
\email{dpc@physics.uvm.edu}
\affiliation{
Department of Physics\\
University of Vermont\\
Burlington, VT 05405-0125}

\date{\today}

\begin{abstract}
Using variational mean-field theory, many-body dissipative effects on the threshold law for quantum sticking and reflection of neutral and charged particles are examined. For the case of an ohmic bosonic bath, we study the effects of the infrared divergence on the probability of sticking and obtain a non-perturbative expression for the sticking rate.  We find that for weak dissipative coupling $\alpha$, the low energy threshold laws for quantum sticking are modified by an infrared singularity in the bath.  The sticking probability for a neutral particle with incident energy $E\to 0$ behaves asymptotically as  
${\it s}\sim E^{(1+\alpha)/2(1-\alpha)}$; for a charged particle, we obtain ${\it s}\sim E^{\alpha/2(1-\alpha)}$.  Thus, ``quantum mirrors'' --surfaces that become perfectly reflective to particles with incident energies asymptotically approaching zero-- can also exist for charged particles.

\end{abstract}

\maketitle
Since the very early years of quantum theory, theorists have considered the interaction of low-energy atoms and molecules with surfaces \cite{lj1,*lj2,*lj3}.   In comparison to a classical particle, a quantum particle at low energy was predicted to have a reduced probability to adsorb to surfaces.  The reason is despite the long-range attractive van der Waals interaction between a neutral particle and surface, at sufficiently low energies, quantum particles have little probability of coming near the surface  \cite{dpc92}.  

This effect is named ``quantum reflection,'' and it is a simple result of the wave-like nature of low-energy particles moving in a finite-ranged attractive potential.  This reduction in the particle's probability density near the surface leads to a reduction in the transition probability of the particle to a state bound to the surface.    In one of the earliest applications of quantum perturbation theory, Lennard-Jones and Devonshire concluded that the probability of a neutral particle with energy $E$ sticking to the surface should vanish as $\sqrt{E}$ as $E\to 0$.  

In contrast, charged particles do not experience the effects of quantum reflection.  Far from the surface, charged particles interact with the surface through a Coulomb potential.  Due to the slow spatial variation of the Coulomb potential, incident particles behave semiclassically.  As a result, Clougherty and Kohn  \cite{dpc92} found that the sticking probability should tend to a non-vanishing constant as $E\to 0$.   

The seemingly universal scaling law for neutral particles was shown to hold even within a non-perturbative model that includes arbitrarily strong quantum fluctuations of the surface  \cite{dpc92, dpc03}.  This model however was regularized 
with the use of a low-frequency cutoff.  Thus the effects of an infrared divergence involving low frequency excitations were not included in the analysis.  

In the eighties, experiments went to sub-milliKelvin temperatures to look for this threshold law scaling in a variety of physical systems without success \cite{doyle91}.  Theorists  \cite{carraro92} realized that the experiments suffered from unwanted interactions from a substrate supporting the target of a superfluid helium film.  By increasing the thickness of the film, the next generation of experiments \cite{yu93}  produced data consistent with the $\sqrt{E}$ law, and the controversy subsided.

In recent years, with dramatic advances in producing and manipulating ultracold atoms, there is renewed interest in interactions between  low-energy atoms and surfaces.  New technologies have been proposed that rely on the quantum dynamics of ultracold atoms near surfaces;  microfabricated devices called ``atom chips'' would store and manipulate cold atoms near surfaces for quantum information processing and precision metrology \cite{chip04}.  Our understanding of device performance will depend in part on our understanding of ultracold atom-surface interactions.  Experiment is now in a position to test detailed theoretical predictions on the behavior of low-energy sticking and scattering from surfaces.  

In this Letter, we consider theoretically a non-perturbative model that focusses on the effects of low-frequency excitations on quantum reflection and sticking.  We follow the mean-field variational method introduced by Silbey and Harris in their analysis of the quantum dynamics of the spin-boson model \cite{sb}.  Using this method we analyze the effects of the infrared divergence on the sticking process.  Our analysis reveals two distinct scaling regimes in the parameter space in analogy with localized and delocalized phases in the spin-boson model.  In the delocalized regime, an infrared divergence in the bath is cutoff by an energy scale that depends on the incident energy of the particle $E$.  As a consequence, we find that both the threshold laws for neutral and charged particles are modified by the dissipative coupling strength $\alpha$.  As a result of the low frequency fluctuations, the threshold law for neutral particles is no longer universal, and the threshold law for charged particles no longer precludes perfect reflection at ultralow energies.


We take for our model a particle coupled to a bath of oscillators
\begin{equation}
H=H_p+H_b+H_c
\end{equation}
where
\begin{eqnarray}
H_p&=&E c_k^\dagger c_k -E_b b^\dagger b,\\
H_b&=&\sum_q{\omega_q {a_q^\dagger} a_q},\\
H_c&=&-i(c_k^\dagger b+b^\dagger c_k)g_{1}\sum_q \sigma\left(\omega_{q}\right) \ ({a_q-a_q^\dagger}) 
-ic_k^\dagger c_k g_{2}\sum_q \sigma\left(\omega_{q}\right)\ ({a_q-a_q^\dagger}) \nonumber\\
&&-i b^\dagger b g_{3}\sum_q \sigma\left(\omega_{q}\right) \ ({a_q-a_q^\dagger}) 
\end{eqnarray}
where $g_{1}$, $g_{2}$ and $g_{3}$ are model coupling constants and $\sigma\left(\omega_{q}\right)$ depends on the specific particle-excitation coupling mechanism. $c_k^\dagger$ ($c_k$) creates (annihilates) a particle in the entrance channel $\left|k\right\rangle$ with energy 
$E$; $b^\dagger$ ($b$) creates (annihilates) a particle in the bound state $\left|b\right\rangle$ with energy
 $-E_b$.  
$a_q^\dagger$ ($a_q$) creates (annihilates) a boson in the target bath with energy $\omega_q$.  (We use natural units throughout where $\hbar=1$.)  We work in the regime where $E\ll E_b$.  We neglect the probability of ``prompt'' inelastic scattering, where bosons are created and the particle escapes to infinity with degraded energy, as the phase space available for these processes vanishes as $E\to 0$.  Thus only the incoming and bound channels are retained for the particle.

We consider a model with ohmic dissipative spectral density. Such a model can be realized with a semi-infinite elastic solid where the incident particle couples to the surface strain.   The spectral density function that characterizes the coupling to the excitation bath is given by
\begin{equation}
J(\omega)\equiv\sum_{q}g^{2}_{1}\sigma^{2}\left(\omega_{q}\right)\delta(\omega-\omega_{q})=g^{2}_{1}\rho\omega
\end{equation} 
where $\rho$ is a frequency-independent constant.

This model differs in an important way from the model of  Ref.~\cite{dpc92} where low frequency modes were cutoff to prevent an infrared divergence in the rms displacement of the surface atom. In this model, low frequency modes are included, and their effects on quantum reflection and sticking are the focus of this  study. 

We start with the variational approach used by Silbey and Harris \cite{sb} for the ohmic spin-boson model.  A generalized unitary transformation $U=e^S$  is first performed on the Hamiltonian $H$, with 
\begin{equation}
S=i b^\dagger b x
\end{equation}
and 
\begin{equation}
x=\sum_q {{f_q\over\omega_q} (a_q+a_q^\dagger)}
\end{equation}
The variational parameters to be determined are denoted by $f_q$. The unitary transformation displaces the oscillators to new equilibrium positions in the presence of the particle bound to the surface and leaves the oscillators unshifted when the particle is in the continuum state. 

The transformed Hamiltonian $\tilde H$ is given by
\begin{eqnarray}
{\tilde H}&=&e^S H e^{-S}\\
&=&{\tilde H_p}+{\tilde H_b}+{\tilde H_c}
\end{eqnarray}
where
\begin{eqnarray}
{\tilde H_p}&=&E c_k^\dagger c_k -{\tilde E_b}b^\dagger b,\\
{\tilde H_c}&=&-ic^{\dagger}_{k}b\sum_{q}g_{1q}(a_{q}-a_{q}^{\dagger})e^{-ix}-ib^{\dagger}c_{k}e^{ix}\sum_{q}g_{1q}(a_{q}-a_{q}^{\dagger})\nonumber\\
&&-ic_{k}^{\dagger}c_{k}\sum_{q}g_{2q}(a_{q}-a_{q}^{\dagger})-ib^{\dagger}b\sum_{q}(g_{3q}-f_{q})(a_{q}-a_{q}^{\dagger})\\
{\tilde H_b}&=&H_b\\
{\tilde E_{b}}&=&E_{b}+\sum_{q}\frac{2f_{q}g_{3q}-f_{q}^{2}}{\omega_{q}}
\end{eqnarray}
and where $g_{iq}\equiv g_{i}\sigma\left(\omega_{q}\right)$. We define a mean transitional matrix element $\Delta$ 
\begin{equation}
\Delta\equiv i\left\langle e^{ix}\sum_{q}g_{1q}(a_{q}-a_{q}^{\dagger})\right\rangle
\end{equation}
where $ \langle\cdots\rangle$ denotes the expectation over the bath modes.

The Hamiltonian is then separated into the following form 
\begin{equation}
{\tilde H}=H_0+V
\end{equation}
where $V$ is chosen such that $\left\langle V\right\rangle=0$.  Hence, we obtain
\begin{eqnarray}
H_0&=&E c_{k}^{\dagger}c_{k}-\tilde{E_{b}}b^{\dagger}b-\Delta^{*}c_{k}^{\dagger}b-\Delta b^{\dagger}c_{k}+\sum_{q}\omega_{q}a_{q}^{\dagger}a_{q}\\
V&=&-c^{\dagger}_{k}b\left(i\sum_{q}g_{1q}(a_{q}-a_{q}^{\dagger})e^{-ix}-\Delta^{*}\right)-b^{\dagger}c_{k}\left(ie^{ix}\sum_{q}g_{1q}(a_{q}-a_{q}^{\dagger})-\Delta\right)\nonumber\\
&&-ic_{k}^{\dagger}c_{k}\sum_{q}g_{2q}(a_{q}-a_{q}^{\dagger})-ib^{\dagger}b\sum_{q}(g_{3q}-f_{q})(a_{q}-a_{q}^{\dagger})
\end{eqnarray}


We calculate the ground state energy of ${H_0}$ in terms of the variational parameters $\left\{f_{q}\right\}$ and minimize to obtain the following condition
\begin{equation}
f_{q}\left(1+\frac{\epsilon+2\Delta^{2}\omega_{q}^{-1}}{\sqrt{\epsilon^{2}+4\Delta^{2}}}\right)=g_{3q}\left(1+\frac{\epsilon}{\sqrt{\epsilon^{2}+4\Delta^{2}}}\right)+\frac{2\Delta\sqrt{u}g_{1q}}{\sqrt{\epsilon^{2}+4\Delta^{2}}}
\label{selffq}
\end{equation}
which is an implicit equation for $f_{q}$. For convenience, in the above we have defined
\begin{equation}
\epsilon=E+\tilde{E_{b}}=E+E_{b}+\sum_{q}\frac{2f_{q}g_{3q}-f_{q}^{2}}{\omega_{q}}
\label{epsilon}
\end{equation}
and
\begin{eqnarray}
\Delta&=&\sqrt{u}\Omega_{1},
\label{delta}\\
u&\equiv&e^{-\sum_{q}\frac{f_{q}^{2}}{\omega_{q}^{2}}},
\label{u1}\\
\Omega_{1}&\equiv&\sum_{q}\frac{g_{1q}f_{q}}{\omega_{q}}
\label{omega1}
\end{eqnarray}\\
Under the condition ${\Delta}\ll{\epsilon}$,
Eq.(\ref{selffq}) can be simplified to
\begin{equation}
f_{q}=\frac{g_{3q}}{1+\frac{z}{\omega_{q}}}
\label{fq}
\end{equation}
where
\begin{equation}
z\equiv\frac{\Delta^{2}}{\epsilon}
\label{defz}
\end{equation}

Using Eq.(\ref{fq}), Eqs.(\ref{u1}), (\ref{omega1}) and (\ref{epsilon}) can be rewritten in terms of $z$
\begin{eqnarray}
u&=&(1+\frac{\omega_{c}}{z})^{-g_{3}^{2}\rho}e^{g_{3}^{2}\rho/(1+\frac{z}{\omega_{c}})}
\label{u2}\\
\Omega_{1}&=&g_{1}g_{3}\rho\omega_{c}-g_{1}g_{3}\rho z \ln\frac{\omega_{c}+z}{z}\\
\epsilon&=&E+E_{b}+g^{2}_{3}\omega_{c}-g^{2}_{3}\rho\omega_{c}\frac{z}{\omega_{c}+z}
\end{eqnarray}
where $\omega_{c}$ is the upper cutoff frequency of the bath. According to Eq.(\ref{defz}) and the condition $\Delta\ll\epsilon$, ${z}\ll{\omega_{c}}$ must be satisfied.  This leads to the following 
\begin{eqnarray}
\Omega_{1}&\approx&g_{1}g_{3}\rho\omega_{c}
\label{solomega1}\\
\epsilon&\approx&E+E_{b}+g_{3}^{2}\rho\omega_{c}
\label{bias}
\end{eqnarray}

Substitution into Eq. (\ref{defz}) gives the self-consistent equation for $z$
\begin{equation}
z=K(1+\frac{\omega_{c}}{z})^{-g_{3}^{2}\rho}e^{g_{3}^{2}\rho/(1+\frac{z}{\omega_{c}})}
\label{selfz}
\end{equation}
where
\begin{equation}
K\approx\frac{(g_{1}g_{3}\rho\omega_{c})^{2}}{E+E_{b}+g_{3}^{2}\rho\omega_{c}}
\end{equation}

It is straightforward to find the following closed-form expression for $z$, valid for $z\ll \omega_c$
\begin{equation}
z\approx K(\frac{eK}{\omega_{c}})^{\frac{\alpha}{1-\alpha}}
\label{z}
\end{equation}
where  $\alpha$, the dissipative coupling strength, is given by  $\alpha\equiv g^{2}_{3}\rho$. 

Depending on the value of $\alpha$, there are two solutions to the variational parameters $f_{q}$.  We see from Eq.~\ref{z} that as $\alpha\to 1$, $z\to 0$. Thus,
\begin{equation}
\label{eq:fq}
f_{q}\approx
\begin{cases}
g_{3q} & \text{$\alpha \ge 1$} \\
\frac{g_{3q}}{1+\frac{z}{\omega_{q}}} & \text{$\alpha < 1$}
\end{cases}
\end{equation}
In the regime where $\alpha <1$, we see that the parameter $f_q$ for excitations whose frequency $\omega_q \ll z$ vanishes as $\omega_q\to 0$.  It is this weakening of the coupling to non-adiabatic excitations that allows us to extract a finite mean transitional matrix element.  In the process, the sticking rate is altered from the perturbative result.

We can now show that the condition ${\Delta}\ll{\epsilon}$ is satisfied so that our variational solution is self-consistent.  According to Eq.~(\ref{defz}),
${\Delta}/{\epsilon}=\sqrt{{z}/{\epsilon}}$.
For $\alpha\geq1$, $z=0$, so ${\Delta}=0$ and ${\Delta}\ll{\epsilon}$ holds true. For $\alpha<1$, $z\sim g^{\frac{2}{1-\alpha}}_{1}$. The coupling constant $g_1$ has a dependence on the initial energy of the particle $E$.  This can be seen from the transition matrix element
\begin{equation}
g_{1q}=-i  \langle b, 1_q|H_c| k, 0\rangle
\label{g1}
\end{equation}
The amplitude of the initial state in the vicinity of the surface is suppressed by quantum reflection.  It is a simple consequence of wave mechanics \cite{dpc92} that in the low energy regime, $g_{1q}\sim \sqrt{E}$ as $E\to 0$ for a neutral particle.    For a charged particle, the coupling constant behaves as $g_{1q}\sim {E^{1/4}}$ as $E\to 0$, as it is not subject to the effects of quantum reflection.
Thus in either case, the mean-field amplitude $\Delta$ becomes arbitrarily small as $E$ tends to zero, while $\epsilon$ approaches a non-zero value.  Consequently the conditions for our variational solution are always satisfied for sufficiently cold particles. 

For ${\Delta}\ll{\epsilon}$, the rate of incoming atoms sticking to the surface can be calculated using Fermi's golden rule \cite{leggett, *leggett2}:
\begin{equation}
R=2\pi \sum_{q}\left|\left\langle b,1_{q}\left|\tilde{H}_{c}\right|k,0\right\rangle\right|^{2}\delta\left(-\tilde{E}_{b}-E+\omega_{q}\right)
\label{R1}
\end{equation}
where $|1_q\rangle$ denotes a state of one excitation with wave vector $q$.

After calculating the relevant matrix elements, the rate becomes
\begin{equation}
R=2\pi e^{-\sum_{q}\frac{f_{q}^{2}}{\omega_{q}^{2}}}\sum_{q}\left(g_{1q}-\frac{f_{q}}{\omega_{q}}\sum_{q^{'}}\frac{f_{q^{'}}g_{1q^{'}}}{\omega_{q^{'}}}\right)^{2}\delta\left(-\tilde{E}_{b}-E+\omega_{q}\right)
\label{r3}
\end{equation}
After some algebra, we find the leading order of the rate $R$ in the incident energy $E$ to be
\begin{equation}
R=2\pi(\frac{z}{\omega_{c}})^{\alpha}e^{\alpha}g^{2}_{1}\rho E_{b}\left(\frac{E_b}{E_{b}+\alpha\omega_{c}}\right)
\label{r2}
\end{equation}
where $z$ given in Eq.~\ref{z} is a constant with a power dependence on $g_{1}$.\\

We compare this rate to that obtained by Fermi's golden rule on the untransformed Hamiltonian
\begin{eqnarray}
R&=&2\pi \sum_{q}\left|\left\langle b,1_{q}\left|H_{c}\right|k,0\right\rangle\right|^{2}\delta\left(-E_{b}-E+\omega_{q}\right)\nonumber\\
&=&2\pi g^{2}_{1}\rho E_{b}
\end{eqnarray}
The matrix elements of transformed Hamiltonian $\tilde{H}_{c}$ are reduced by a Franck-Condon factor which gives the non-perturbative rate with an additional dependence on $z$. 

The coupling constant $g_{1}$ can be expressed  in terms of a matrix element of the unperturbed states using Eq.~\ref{g1}.  We take $H_{c}$ to have the general form in coordinate space
\begin{equation}
H_{c}=-i U(x)\sum_{q}\sigma\left(\omega_{q}\right)\left(a_{q}-a^{\dagger}_{q}\right)
\end{equation}
The coupling constant $g_{1}$ is given by 
\begin{equation}
g_{1}=\left\langle k\left|U\right|b\right\rangle=\int^{\infty}_{0}{\phi^{*}_{k}(x)U(x)\phi_{b}(x)dx}
\label{g1k}
\end{equation}
(We have assumed the case of normal incidence, however results for the more general case follow from decomposing the wave vector into normal and transverse components \cite{berlinsky}.) 

 The continuum wave functions have the asymptotic form for a neutral particle
\begin{equation}
\phi_{k}(x)\stackrel{k\rightarrow0}{\sim} {k\ h_1(x)}
\label{phikn}
\end{equation}
and for a charged particle \cite{dpc92},
\begin{equation}
\phi_{k}(x)\stackrel{k\rightarrow0}{\sim} {\sqrt{k}\ h_2(x)}
\label{phikc}
\end{equation}
where $k=\sqrt{2mE}$, and $h_i(x)$ are functions, independent of $E$. 

The probability of sticking to the surface $\it s$ is the sticking rate per surface area per unit incoming particle flux. Hence,
${\it s}(E)=\sqrt{2\pi^2 m\over { E}}R$. (We use delta-function normalization for the continuum wave functions.) From Eq.~\ref{r2} we conclude that with $\alpha<1$ for a neutral particle,
\begin{equation}
{\it s}(E)\sim C_1 {E^{(1+\alpha)/2(1-\alpha)}},\ \ \ E\to 0
\end{equation}
and for a charged particle,
\begin{equation}
{\it s}(E)\sim C_2 {E^{\alpha/2(1-\alpha)}},\ \ \ E\to 0
\end{equation}
where $C_i$ are energy-independent constants.
  
In summary, we have considered the effects of the infrared singularity resulting from interaction with an ohmic bath on surface sticking.  We calculated using a variational mean-field method the sticking rate as a function of the incident energy in the low-energy asymptotic regime.   We have shown  that for an ohmic excitation bath the threshold rate for neutral particles decreases more rapidly with decreasing energy $E$, in comparison with the perturbative rate.   We predict a new threshold law for neutral particle surface sticking, where the energy dependence depends on the dissipative coupling $\alpha$.  

The new threshold law is a result of a bosonic orthogonality catastrophe \cite{mahan}; the ground states of the bath with different particle states are orthogonal.  The sticking transition amplitude acquires a Franck-Condon factor whose infrared singularity is cutoff by $z$.  As with the x-ray absorption edge \cite{mahan}, a new power law results at threshold.  The low-frequency fluctuations alter the power law to a bath-dependent non-universal exponent.  

For the case of charged particles, we find that dissipative coupling causes the sticking probability to vanish as $E\to 0$, in contrast to the perturbative result \cite{dpc92}.  Thus, ``quantum mirrors'' --surfaces that become perfectly reflective to particles with incident energies asymptotically approaching zero-- can also exist for charged particles.

\begin{acknowledgments}
DPC thanks Daniel Fisher, Walter Kohn and James Langer for stimulating discussions on various aspects of this problem.  We gratefully acknowledge support by the National Science Foundation  (DMR-0814377). 
\end{acknowledgments}
 
\bibliography{qs}

\begin{thebibliography}{14}%
\makeatletter
\providecommand \@ifxundefined [1]{%
 \@ifx{#1\undefined}
}%
\providecommand \@ifnum [1]{%
 \ifnum #1\expandafter \@firstoftwo
 \else \expandafter \@secondoftwo
 \fi
}%
\providecommand \@ifx [1]{%
 \ifx #1\expandafter \@firstoftwo
 \else \expandafter \@secondoftwo
 \fi
}%
\providecommand \natexlab [1]{#1}%
\providecommand \enquote  [1]{``#1''}%
\providecommand \bibnamefont  [1]{#1}%
\providecommand \bibfnamefont [1]{#1}%
\providecommand \citenamefont [1]{#1}%
\providecommand \href@noop [0]{\@secondoftwo}%
\providecommand \href [0]{\begingroup \@sanitize@url \@href}%
\providecommand \@href[1]{\@@startlink{#1}\@@href}%
\providecommand \@@href[1]{\endgroup#1\@@endlink}%
\providecommand \@sanitize@url [0]{\catcode `\\12\catcode `\$12\catcode
  `\&12\catcode `\#12\catcode `\^12\catcode `\_12\catcode `\%12\relax}%
\providecommand \@@startlink[1]{}%
\providecommand \@@endlink[0]{}%
\providecommand \url  [0]{\begingroup\@sanitize@url \@url }%
\providecommand \@url [1]{\endgroup\@href {#1}{\urlprefix }}%
\providecommand \urlprefix  [0]{URL }%
\providecommand \Eprint [0]{\href }%
\providecommand \doibase [0]{http://dx.doi.org/}%
\providecommand \selectlanguage [0]{\@gobble}%
\providecommand \bibinfo  [0]{\@secondoftwo}%
\providecommand \bibfield  [0]{\@secondoftwo}%
\providecommand \translation [1]{[#1]}%
\providecommand \BibitemOpen [0]{}%
\providecommand \bibitemStop [0]{}%
\providecommand \bibitemNoStop [0]{.\EOS\space}%
\providecommand \EOS [0]{\spacefactor3000\relax}%
\providecommand \BibitemShut  [1]{\csname bibitem#1\endcsname}%
\let\auto@bib@innerbib\@empty
\bibitem [{\citenamefont {Lennard-Jones}\ and\ \citenamefont
  {Strachan}(1935)}]{lj1}%
  \BibitemOpen
  \bibfield  {author} {\bibinfo {author} {\bibfnamefont {J.~E.}\ \bibnamefont
  {Lennard-Jones}}\ and\ \bibinfo {author} {\bibfnamefont {C.}~\bibnamefont
  {Strachan}},\ }\href@noop {} {\bibfield  {journal} {\bibinfo  {journal}
  {Proc.\ R.\ Soc.\ London, Ser. A}\ }\textbf {\bibinfo {volume} {150}},\
  \bibinfo {pages} {442} (\bibinfo {year} {1935})}\BibitemShut {NoStop}%
\bibitem [{\citenamefont {Lennard-Jones}\ and\ \citenamefont
  {Devonshire}(1936{\natexlab{a}})}]{lj2}%
  \BibitemOpen
  \bibfield  {author} {\bibinfo {author} {\bibfnamefont {J.~E.}\ \bibnamefont
  {Lennard-Jones}}\ and\ \bibinfo {author} {\bibfnamefont {A.~F.}\ \bibnamefont
  {Devonshire}},\ }\href@noop {} {\bibfield  {journal} {\bibinfo  {journal}
  {Proc. R. Soc. London, Ser. A}\ }\textbf {\bibinfo {volume} {156}},\ \bibinfo
  {pages} {6} (\bibinfo {year} {1936}{\natexlab{a}})}\BibitemShut {NoStop}%
\bibitem [{\citenamefont {Lennard-Jones}\ and\ \citenamefont
  {Devonshire}(1936{\natexlab{b}})}]{lj3}%
  \BibitemOpen
  \bibfield  {author} {\bibinfo {author} {\bibfnamefont {J.~E.}\ \bibnamefont
  {Lennard-Jones}}\ and\ \bibinfo {author} {\bibfnamefont {A.~F.}\ \bibnamefont
  {Devonshire}},\ }\href@noop {} {\bibfield  {journal} {\bibinfo  {journal}
  {Proc. R. Soc. London, Ser. A}\ }\textbf {\bibinfo {volume} {156}},\ \bibinfo
  {pages} {29} (\bibinfo {year} {1936}{\natexlab{b}})}\BibitemShut {NoStop}%
\bibitem [{\citenamefont {Clougherty}\ and\ \citenamefont
  {Kohn}(1992)}]{dpc92}%
  \BibitemOpen
  \bibfield  {author} {\bibinfo {author} {\bibfnamefont {D.~P.}\ \bibnamefont
  {Clougherty}}\ and\ \bibinfo {author} {\bibfnamefont {W.}~\bibnamefont
  {Kohn}},\ }\href@noop {} {\bibfield  {journal} {\bibinfo  {journal} {Phys.\
  Rev.\ B}\ }\textbf {\bibinfo {volume} {46}},\ \bibinfo {pages} {4921}
  (\bibinfo {year} {1992})}\BibitemShut {NoStop}%
\bibitem [{\citenamefont {Clougherty}(2003)}]{dpc03}%
  \BibitemOpen
  \bibfield  {author} {\bibinfo {author} {\bibfnamefont {D.~P.}\ \bibnamefont
  {Clougherty}},\ }\href@noop {} {\bibfield  {journal} {\bibinfo  {journal}
  {Phys.\ Rev.\ Lett.}\ }\textbf {\bibinfo {volume} {91}},\ \bibinfo {pages}
  {226105} (\bibinfo {year} {2003})}\BibitemShut {NoStop}%
\bibitem [{\citenamefont {Doyle}\ \emph {et~al.}(1991)\citenamefont {Doyle},
  \citenamefont {Sandberg}, \citenamefont {Yu}, \citenamefont {Cesar},
  \citenamefont {Kleppner},\ and\ \citenamefont {Greytak}}]{doyle91}%
  \BibitemOpen
  \bibfield  {author} {\bibinfo {author} {\bibfnamefont {J.~M.}\ \bibnamefont
  {Doyle}}, \bibinfo {author} {\bibfnamefont {J.~C.}\ \bibnamefont {Sandberg}},
  \bibinfo {author} {\bibfnamefont {I.~A.}\ \bibnamefont {Yu}}, \bibinfo
  {author} {\bibfnamefont {C.~L.}\ \bibnamefont {Cesar}}, \bibinfo {author}
  {\bibfnamefont {D.}~\bibnamefont {Kleppner}}, \ and\ \bibinfo {author}
  {\bibfnamefont {T.~J.}\ \bibnamefont {Greytak}},\ }\href@noop {} {\bibfield
  {journal} {\bibinfo  {journal} {Phys. Rev. Lett.}\ }\textbf {\bibinfo
  {volume} {67}},\ \bibinfo {pages} {603} (\bibinfo {year} {1991})}\BibitemShut
  {NoStop}%
\bibitem [{\citenamefont {Carraro}\ and\ \citenamefont
  {Cole}(1992)}]{carraro92}%
  \BibitemOpen
  \bibfield  {author} {\bibinfo {author} {\bibfnamefont {C.}~\bibnamefont
  {Carraro}}\ and\ \bibinfo {author} {\bibfnamefont {M.~W.}\ \bibnamefont
  {Cole}},\ }\href@noop {} {\bibfield  {journal} {\bibinfo  {journal} {Phys.\
  Rev.\ Lett.}\ }\textbf {\bibinfo {volume} {68}},\ \bibinfo {pages} {412}
  (\bibinfo {year} {1992})}\BibitemShut {NoStop}%
\bibitem [{\citenamefont {Yu}\ \emph {et~al.}(1993)\citenamefont {Yu},
  \citenamefont {Doyle}, \citenamefont {Sandberg}, \citenamefont {Cesar},
  \citenamefont {Kleppner},\ and\ \citenamefont {Greytak}}]{yu93}%
  \BibitemOpen
  \bibfield  {author} {\bibinfo {author} {\bibfnamefont {I.~A.}\ \bibnamefont
  {Yu}}, \bibinfo {author} {\bibfnamefont {J.~M.}\ \bibnamefont {Doyle}},
  \bibinfo {author} {\bibfnamefont {J.~C.}\ \bibnamefont {Sandberg}}, \bibinfo
  {author} {\bibfnamefont {C.~L.}\ \bibnamefont {Cesar}}, \bibinfo {author}
  {\bibfnamefont {D.}~\bibnamefont {Kleppner}}, \ and\ \bibinfo {author}
  {\bibfnamefont {T.~J.}\ \bibnamefont {Greytak}},\ }\href@noop {} {\bibfield
  {journal} {\bibinfo  {journal} {Phys. Rev. Lett.}\ }\textbf {\bibinfo
  {volume} {71}},\ \bibinfo {pages} {1589} (\bibinfo {year}
  {1993})}\BibitemShut {NoStop}%
\bibitem [{\citenamefont {Groth}\ \emph {et~al.}(2004)\citenamefont {Groth},
  \citenamefont {Kr{\"u}ger}, \citenamefont {Wildermuth}, \citenamefont
  {Folman}, \citenamefont {Fernholz}, \citenamefont {Mahalu}, \citenamefont
  {Bar-Joseph},\ and\ \citenamefont {Schmiedmayer}}]{chip04}%
  \BibitemOpen
  \bibfield  {author} {\bibinfo {author} {\bibfnamefont {S.}~\bibnamefont
  {Groth}}, \bibinfo {author} {\bibfnamefont {P.}~\bibnamefont {Kr{\"u}ger}},
  \bibinfo {author} {\bibfnamefont {S.}~\bibnamefont {Wildermuth}}, \bibinfo
  {author} {\bibfnamefont {R.}~\bibnamefont {Folman}}, \bibinfo {author}
  {\bibfnamefont {T.}~\bibnamefont {Fernholz}}, \bibinfo {author}
  {\bibfnamefont {D.}~\bibnamefont {Mahalu}}, \bibinfo {author} {\bibfnamefont
  {I.}~\bibnamefont {Bar-Joseph}}, \ and\ \bibinfo {author} {\bibfnamefont
  {J.}~\bibnamefont {Schmiedmayer}},\ }\href@noop {} {\bibfield  {journal}
  {\bibinfo  {journal} {Appl.\ Phys.\ Lett.}\ }\textbf {\bibinfo {volume}
  {85}},\ \bibinfo {pages} {14} (\bibinfo {year} {2004})}\BibitemShut {NoStop}%
\bibitem [{\citenamefont {Silbey}\ and\ \citenamefont {Harris.}(1984)}]{sb}%
  \BibitemOpen
  \bibfield  {author} {\bibinfo {author} {\bibfnamefont {R.}~\bibnamefont
  {Silbey}}\ and\ \bibinfo {author} {\bibfnamefont {R.~A.}\ \bibnamefont
  {Harris.}},\ }\href@noop {} {\bibfield  {journal} {\bibinfo  {journal}
  {J.~Chem.~Phys.}\ }\textbf {\bibinfo {volume} {80}},\ \bibinfo {pages} {2615}
  (\bibinfo {year} {1984})}\BibitemShut {NoStop}%
\bibitem [{\citenamefont {Leggett}\ \emph {et~al.}(1987)\citenamefont
  {Leggett}, \citenamefont {Chakravarty}, \citenamefont {Dorsey}, \citenamefont
  {Fisher},\ and\ \citenamefont {Garg}}]{leggett}%
  \BibitemOpen
  \bibfield  {author} {\bibinfo {author} {\bibfnamefont {A.~J.}\ \bibnamefont
  {Leggett}}, \bibinfo {author} {\bibfnamefont {S.}~\bibnamefont
  {Chakravarty}}, \bibinfo {author} {\bibfnamefont {A.~T.}\ \bibnamefont
  {Dorsey}}, \bibinfo {author} {\bibfnamefont {M.~P.~A.}\ \bibnamefont
  {Fisher}}, \ and\ \bibinfo {author} {\bibfnamefont {A.}~\bibnamefont
  {Garg}},\ }\href@noop {} {\bibfield  {journal} {\bibinfo  {journal}
  {Rev.~Mod.~Phys.}\ }\textbf {\bibinfo {volume} {59}},\ \bibinfo {pages} {1}
  (\bibinfo {year} {1987})}\BibitemShut {NoStop}%
\bibitem [{\citenamefont {Leggett}\ \emph {et~al.}(1995)\citenamefont
  {Leggett}, \citenamefont {Chakravarty}, \citenamefont {Dorsey}, \citenamefont
  {Fisher},\ and\ \citenamefont {Garg}}]{leggett2}%
  \BibitemOpen
  \bibfield  {author} {\bibinfo {author} {\bibfnamefont {A.~J.}\ \bibnamefont
  {Leggett}}, \bibinfo {author} {\bibfnamefont {S.}~\bibnamefont
  {Chakravarty}}, \bibinfo {author} {\bibfnamefont {A.~T.}\ \bibnamefont
  {Dorsey}}, \bibinfo {author} {\bibfnamefont {M.~P.~A.}\ \bibnamefont
  {Fisher}}, \ and\ \bibinfo {author} {\bibfnamefont {A.}~\bibnamefont
  {Garg}},\ }\href@noop {} {\bibfield  {journal} {\bibinfo  {journal}
  {Rev.~Mod.~Phys.}\ }\textbf {\bibinfo {volume} {67}},\ \bibinfo {pages} {725}
  (\bibinfo {year} {1995})}\BibitemShut {NoStop}%
\bibitem [{\citenamefont {Zimmerman}\ and\ \citenamefont
  {Berlinsky}(1983)}]{berlinsky}%
  \BibitemOpen
  \bibfield  {author} {\bibinfo {author} {\bibfnamefont {D.~S.}\ \bibnamefont
  {Zimmerman}}\ and\ \bibinfo {author} {\bibfnamefont {A.~J.}\ \bibnamefont
  {Berlinsky}},\ }\href@noop {} {\bibfield  {journal} {\bibinfo  {journal}
  {Can. J. Phys.}\ }\textbf {\bibinfo {volume} {61}},\ \bibinfo {pages} {50}
  (\bibinfo {year} {1983})}\BibitemShut {NoStop}%
\bibitem [{\citenamefont {Mahan}(1981)}]{mahan}%
  \BibitemOpen
  \bibfield  {author} {\bibinfo {author} {\bibfnamefont {G.~D.}\ \bibnamefont
  {Mahan}},\ }\href@noop {} {\emph {\bibinfo {title} {Many-Particle Physics}}}\
  (\bibinfo  {publisher} {Plenum Press},\ \bibinfo {address} {New York},\
  \bibinfo {year} {1981})\ p.\ \bibinfo {pages} {761}\BibitemShut {NoStop}%
\end{thebibliography}%
\end{document}